\documentclass[a4paper,11pt]{article}
\usepackage{aaskaiid}
\usepackage{orcidlink}

\newcommand{\HI}{H\,\textsc{i}}

\title{The Road to Identifying the Earliest Radio-Powerful AGN with the SKA}
\ShortTitle{The Earliest Radio AGN}

\author[1,2]{Jose Afonso\orcidlink{0000-0002-9149-2973}}
\ShortName{Afonso et al.} 
\author[1,2,3]{Stergios Amarantidis\orcidlink{0000-0001-7948-5714}}
\author[4]{Stas Shabala\orcidlink{0000-0001-5064-0493}}
\author[4]{Ross J. Turner\orcidlink{0000-0002-4376-5455}}
\author[5]{Luca Ighina\orcidlink{0000-0003-1516-9450}}
\author[6,7]{Mojtaba Raouf\orcidlink{0000-0002-1496-3591}}
\author[1,2]{Nuno Covas}
\author[1,2,8]{Pedro Martins}
\author[9]{Nick Seymour\orcidlink{0000-0003-3506-5536}}
\author[10]{Alessandro Caccianiga\orcidlink{0000-0002-2339-8264}}
\author[9]{Alexander Hedge\orcidlink{0009-0007-5594-6476}}
\author[11]{Jess W. Broderick\orcidlink{0000-0002-2239-6099}}
\author[1,2]{Davi Barbosa}
\author[12]{Isabella Prandoni\orcidlink{0000-0001-9680-7092}}
\author[13]{Sabyasachi Pal\orcidlink{0000-0003-2325-8509}}
\author[1,2]{Bruno Arsioli\orcidlink{0000-0001-8166-6602}}
\author[1,2]{Luis Barroso}
\author[1,2]{Rodrigo Carvajal\orcidlink{0000-0002-0545-1113}}
\author[1,2]{Jo\~{a}o Tiago}
\author[14]{Andrew Hopkins\orcidlink{0000-0002-6097-2747}}
\author[15]{Manuela Magliocchetti\orcidlink{0000-0001-9158-4838}}
\author[1,2]{Israel Matute\orcidlink{0000-0003-1177-3896}}
\author[1,2]{Ciro Pappalardo\orcidlink{0000-0003-2606-6019}}

\affiliation[1]{Instituto de Astrofísica e Ciências do Espaço, Portugal}
\affiliation[2]{Faculdade de Ciências da Universidade de Lisboa, Portugal}
\emailAdd{jmafonso@ciencias.ulisboa.pt}
\affiliation[3]{Instituto de Radioastronomía Milimétrica (IRAM), Spain}
\affiliation[4]{School of Natural Sciences, University of Tasmania, Australia}
\affiliation[5]{Center for Astrophysics | Harvard \& Smithsonian, USA}
\affiliation[6]{School of Astronomy, Institute for Research in Fundamental Sciences (IPM), Tehran, 19395-5746, Iran}
\affiliation[7]{Leiden Observatory, Leiden University, P.O. Box 9513, 2300 RA Leiden, Netherlands}
\affiliation[8]{Associação RAEGE, Santa Maria, Açores, Portugal}
\affiliation[9]{International Centre for Radio Astronomy Research, Curtin University, Australia}
\affiliation[10]{INAF, Osservatorio Astronomico di Brera, Italy}
\affiliation[11]{SKA Observatory, Science Operations Centre, CSIRO ARRC, 26 Dick Perry Avenue, Kensington, WA 6151, Australia}
\affiliation[12]{INAF - Institute of Radio Astronomy, Italy}
\affiliation[13]{Midnapore City College, India}
\affiliation[14]{School of Mathematical and Physical Sciences, Macquarie University, Australia}
\affiliation[15]{INAF-IAPS, Italy}

\abstract{ 
The Epoch of Reionization (EoR) is one of the most pivotal frontiers in modern
astrophysics, marking the emergence of the first galaxies, stars, and supermassive black holes (SMBHs). Despite unprecedented insights from the Atacama Large Millimetre/submillimetre Array (ALMA) and the James Webb Space Telescope (JWST), we still struggle to explain how $\sim10^{9}\,\rm M_\odot$ SMBHs powering luminous active galactic nuclei (AGN) already exist by $z\sim7$. The recent discovery of powerful radio emission from some of these early AGN is particularly significant, offering new constraints on early black-hole accretion and, with the Square Kilometre Array Observatory (SKAO), the prospect of directly probing neutral hydrogen through 21-cm absorption studies.

Yet progress remains slow: only a few radio-powerful AGN are known at $z>6$, far fewer than theoretical predictions suggest, raising fundamental questions about whether early SMBHs are intrinsically radio-powerful or whether current strategies are hindered by selection biases and incomplete spectral information.

In this chapter we synthesise predictions from state-of-the-art hydrodynamical and semi-analytic simulations with observational constraints from SKAO pathfinder facilities. These models suggest the existence of a substantial, still-undetected population of radio-powerful AGN in the EoR, but show clearly that present surveys are limited by selection biases and incomplete radio spectral information.

We discuss a physically motivated strategy for identifying high-redshift
radio AGN, based on broadband radio spectral energy distributions (SEDs), spectral curvature, dynamical jet evolution, and radio-only redshift estimation, offering a transformative alternative to traditional empirical approaches. 

Finally, we justify how the sensitivity, spectral coverage, and spatial resolution of the SKAO will allow fine-frequency sampling across the 50\,MHz -- 15\,GHz range, revolutionising our ability to identify the earliest radio-powerful AGN and study the birth and growth of SMBHs in the early Universe.
}

\begin{document}
\maketitle

\newcommand{\actaa}{Acta Astron.} 
\newcommand{\araa}{ARA\&A} 
\newcommand{\aar}{A\&ARv} 
\newcommand{\aapr}{A\&ARv} 
\newcommand{\ab}{Astrobiol.} 
\newcommand{\aj}{AJ} 
\newcommand{\apj}{ApJ} 
\newcommand{\apjl}{ApJL} 
\newcommand{\apjs}{ApJSS} 
\newcommand{\ao}{Appl. Opt.} 
\newcommand{\apss}{Astro. \& Space Sci.} 
\newcommand{\aap}{A\&A} 
\newcommand{\aaps}{A\&AS.} 
\newcommand{\baas}{Bull. Am. Astron. Soc.} 
\newcommand{\caa}{Chinese A\&A} 
\newcommand{\cjaa}{Chinese J. A\&A} 
\newcommand{\cqg}{Class. Quantum Gravity} 
\newcommand{\gal}{Galaxies} 
\newcommand{\gca}{Geo. Cosmo. Acta} 
\newcommand{\icarus}{Icarus} 
\newcommand{\jcap}{JCAP} 
\newcommand{\jgr}{J. Geophys. Res.} 
\newcommand{\jgrp}{J. Geophys. Res. Planets} 
\newcommand{\jqsrt}{J. Quant. Spectrosc. Radiat. Transf.} 
\newcommand{\memsai}{Mem. SAIt} 
\newcommand{\mnras}{MNRAS} 
\newcommand{\nat}{Nature} 
\newcommand{\nastro}{Nat. Astron.} 
\newcommand{\ncomms}{Nat. Commun.} 
\newcommand{\nphys}{Nat. Phys.} 
\newcommand{\na}{New Astron.} 
\newcommand{\nar}{New Astron. Rev.} 
\newcommand{\physrep}{Phys. Rep.} 
\newcommand{\pra}{Phys. Rev. A} 
\newcommand{\prb}{Phys. Rev. B} 
\newcommand{\prc}{Phys. Rev. C} 
\newcommand{\prd}{Phys. Rev. D} 
\newcommand{\pre}{Phys. Rev. E} 
\newcommand{\prx}{Phys. Rev. X} 
\newcommand{\prl}{Phys. Rev. Let.} 
\newcommand{\psj}{Planet. Sci. J.} 
\newcommand{\planss}{Planet. Space Sci.} 
\newcommand{\pnas}{Proc. Natl Acad. Sci. USA} 
\newcommand{\procspie}{Proc. SPIE} 
\newcommand{\pasa}{PASA} 
\newcommand{\pasj}{PASJ} 
\newcommand{\pasp}{PASP} 
\newcommand{\rmxaa}{RMXAA} 
\newcommand{\sci}{Science} 
\newcommand{\sciadv}{Sci. Adv.} 
\newcommand{\solphys}{Sol. Phys.} 
\newcommand{\sovast}{Soviet Ast.} 
\newcommand{\ssr}{Space Sci. Rev.} 
\newcommand{\uni}{Universe} 

\section{Introduction} \label{sec:Introduction}

The Epoch of Reionization (EoR) represents one of the most exciting and challenging frontiers of modern astrophysics. It corresponds to the period when the first galaxies, stars, and supermassive black holes (SMBHs) transformed the Universe from an almost completely neutral state, at $z \sim 15-20$, to a nearly fully ionised intergalactic medium (IGM) by $z \sim 6{-}7$ \citep[e.g.,][]{BarkanaLoeb2001,Zaroubi2013}. Understanding the timing, drivers, and processes of this transition is essential to constrain the earliest phases of structure formation and the physical mechanisms that link black hole growth to galaxy assembly.

Over the last two decades, ground-based optical and near-infrared (NIR) surveys have revealed a population of luminous quasars powered by SMBHs with masses $\gtrsim 10^8{-}10^9\,\rm M_\odot$ already in place by $z \sim 7$ \citep[][and references therein]{Fan2006,Mortlock2011,Banados2018,Wang2021,Fan2023}.  These results demonstrate that black holes can grow to billion-solar-mass scales within the first Gyr, posing a major challenge for seeding and growth models.

More recently, the James Webb Space Telescope (JWST) has dramatically extended this frontier. Deep Near-Infrared Camera (NIRCam) and Near-Infrared Spectrograph (NIRSpec) observations have uncovered unexpectedly abundant populations of galaxies out to $z \gtrsim 14$ \citep{Naidu2022,Finkelstein2022,Curtis-Lake2023,Donnan2023,Robertson2023,Carniani2024} and, perhaps even more strikingly, revealed numerous actively accreting black holes with masses of only $10^{6{-}7}\,\rm M_\odot$ at $z \sim 5{-}9$ \citep{Kocevski2023,Harikane2023,Goulding2023,Larson2023,Furtak2024,Maiolino2024,Taylor2025,Juodzbalis2025}. Like their more massive counterparts, these low-mass SMBHs are also difficult to reconcile with standard seeding and growth scenarios, implying either more efficient accretion or heavier initial seeds. At the same time, the Atacama Large Millimetre/submillimetre Array (ALMA) has played a complementary role, providing spectroscopic confirmation and detailed characterisation of the dust and gas content of galaxies and quasars at $z \sim 6{-}9$ through detections of the [C\,\textsc{ii}] 158\,$\mu$m and CO lines, as well as continuum emission  \citep{Inoue2016,Hashimoto2018,Venemans2018,Decarli2018,Wang2019b,Carniani2025,Schouws2025}. 


Taken together, these discoveries are rapidly transforming our view of black-hole formation and early AGN activity. They reveal that SMBH growth and feedback were already widespread and complex within the first few hundred million years, revealing a very early onset of the AGN era and reshaping our understanding of how the first massive black holes emerged and co-evolved with their host galaxies.

Radio observations provide a unique and complementary perspective on this problem. Accretion onto SMBHs can produce powerful relativistic jets, leading to luminous radio emission that directly probes black hole spin, accretion mode, and jet efficiency \citep{HeckmanBest2014,Hardcastle2020}. Radio-powerful AGN (RAGN\footnote{We adopt a more radio-centric nomenclature here, and avoid the usual Radio-Loud vs. Radio-Quiet dichotomy, which historically links to the relation between radio (AGN) and optical (host) emission.}) in the EoR are therefore ideal laboratories to study the coupling between black hole growth and galaxy evolution at the earliest epochs. Equally important, bright RAGN can act as background sources for 21\,cm absorption studies of neutral hydrogen, enabling direct measurements of the ionisation state of the IGM through the so-called \HI\ forest \citep{Carilli2004,Khatri2010,Soltinsky2021,Soltinsky2025}. Realising this potential is one of the most extraordinary science goals of the Square Kilometre Array Observatory (SKAO).

Until very recently, however, progress in identifying RAGN at the highest redshifts was extremely slow. After the discovery of TN\,J0924$-$2201 at $z=5.19$ \citep{vanBreugel1999}, no comparably distant radio-selected sources were found for nearly two decades. Most $z>5$ RAGN revealed during this period were first identified as quasi-stellar objects (QSOs) in optical or NIR surveys and only later shown to host powerful radio emission. It is only with the advent of wide, multi-frequency surveys from SKAO precursors and pathfinders that genuinely radio-selected candidates have begun to emerge at $z>5.5$ -- e.g., the powerful sources at $z=5.55$
\citep{Drouart2020}, $z=6.44$ \citep{Ighina2021}, $z=6.82$ \citep{Banados2021} and even the likely blazar J0410$-$0139 at $z=7.0$ \citep{Banados2025}, hinting at a much larger population still to be uncovered.

The detection of these systems highlights both the opportunities and the challenges of radio-based selection. Historically, the most widely used technique to identify high-redshift radio galaxies has been the ultra-steep spectrum (USS) criterion \citep[e.g.,][]{Roettgering1994,DeBreuck2000a}, motivated by the empirical correlation between spectral index and redshift. While USS selection has proven effective at mJy levels, recent deep surveys in the sub-mJy regime have shown its limitations. Analyses in the COSMOS field demonstrate that faint USS sources are dominated by low-to-intermediate redshift star-forming galaxies, with only a small fraction lying at
$z>2$ (Barbosa et al., submitted). At $\mu$Jy flux densities, spectral index alone fails to discriminate between high-redshift RAGN and the broader faint radio population, underscoring the need for refined, multi-parameter approaches. As an example, GLEAM\,J0856+0223 was identified not through a simple USS cut, but via a low-frequency broadband radio spectral energy distribution (SED) selection that combined spectral index and curvature information, derived from the wide 70–230 MHz bandwidth of the Murchison Widefield Array (MWA), with a compactness criterion (angular size $<5^{\prime\prime}$) and a faint or non-detection in the near-IR, allowing a candidate high-redshift radio source to be identified. ALMA spectroscopy then confirmed its nature as a powerful RAGN at $z=5.5$, with radio luminosity $L_{5\,\mathrm{GHz}} \sim 2\times10^{27}$\,W\,Hz$^{-1}$ \citep{Drouart2020}. This source, among the most luminous RAGN ever found at $z>5$, provides proof-of-concept that efficient selection strategies can reveal the rarest and most luminous systems at the EoR.

Cosmological models of galaxy formation and evolution, such as \textsc{Illustris/TNG} \citep{Vogelsberger2014,Nelson2019}, \textsc{EAGLE} \citep{Schaye2015}, \textsc{Horizon-AGN} \citep{Dubois2014}, \textsc{SIMBA} \citep{Dave2019}, \textsc{GALFORM} \citep{Lacey2016}, and \textsc{Shark} \citep{Lagos2018}, provide a powerful means of anticipating what the next generation of radio surveys may uncover in the early Universe.  They not only allow us to explore the expected abundance and properties of active black holes at very high redshift, but can also guide the development of more efficient selection techniques for identifying RAGN in the wide-area radio surveys currently being conducted with SKAO precursors and pathfinders and, soon, at unprecedented sensitivity and resolution with the SKAO itself.  However, even though most modern cosmological models follow the growth of SMBHs from very early epochs, they face important limitations: in addition to the intrinsic trade-offs between resolution and volume, processes such as SMBH accretion and AGN feedback are modelled following a subgrid approach (simplified phenomenological descriptions of physical processes) that generally lack prescriptions for the radio emission associated with accretion and jet activity. Recent efforts have begun to address this gap by incorporating radio-emission models directly into simulations or by post-processing SMBH properties to predict their radio output. While these operate in a largely uncharted parameter space for black-hole growth in the EoR, they already suggest that a significant population of RAGN could exist at $z>6$, awaiting discovery in the deepest forthcoming radio surveys.  These developments, and their implications for the detectability of the earliest radio AGN, are discussed in Section~\ref{sec:theory}.

The SKAO, its precursors and pathfinders (including the Australian Square Kilometre Array Pathfinder, ASKAP; MeerKAT; MWA; and the Low-Frequency Array, LOFAR) are already transforming this landscape.  Surveys such as the Evolutionary Map of the Universe \citep[EMU;][]{Hopkins2025}, the LOFAR Two-metre Sky Survey \citep[LoTSS;][]{Shimwell2017}, MIGHTEE \citep{Jarvis2016}, MeerKLASS \citep{Santos2016}, and GLEAM \citep{Wayth2015} are unveiling faint radio populations with unprecedented depth and sky coverage, testing selection criteria, and providing the multi-frequency coverage required to characterise radio spectral energy distributions robustly.  Early results demonstrate both the promise and the pitfalls of different approaches: while spectral-curvature selection, as applied to GLEAM sources, appears highly efficient at identifying extreme RAGN \citep{Drouart2020}, simple USS filters at $\mu$Jy depths introduce strong biases (Barbosa et al., submitted).  Some of these developments and their diagnostic potential are discussed further in Section~\ref{sec:observations}.

These lessons are crucial for guiding the design of SKAO surveys and the optimisation of strategies to identify reliable high-redshift candidates, as outlined in Section~\ref{sec:optimisation}.  The experience gained with SKAO precursors and pathfinders is already shaping our understanding of what can realistically be achieved in terms of survey depth, frequency coverage, and source characterisation.  It also highlights the need for consistent multi-wavelength strategies and for robust diagnostics capable of distinguishing genuinely distant RAGN from the overwhelming populations of low- and intermediate-redshift star-forming galaxies.  As the SKAO moves towards full operation, these empirical insights will be essential to maximise the scientific return of its early surveys. 

This chapter aims to highlight some of the scientific and technical challenges that the SKA telescopes will face in exploring the EoR through the detection of RAGN, and to contribute to the preparation for an undoubtedly transformative period in astronomy.  The unprecedented sensitivity, resolution, and spectral coverage of the SKAO will open a new window on the formation of the first massive black holes and on the role of AGN in shaping the earliest galaxies.

\section{Theoretical Prediction for Radio-Powerful AGN in the EoR} \label{sec:theory}

The difficulty in detecting powerful radio galaxies beyond $z\!\sim\!5$ has long fuelled the belief that such systems may be intrinsically rare or even non-existent at very high redshift.  Analyses of flux-limited samples such as the revised Third Cambridge Catalogue of Radio Sources (3CRR) and its extensions \citep[e.g.,][]{Laing1983,Willott2001,Rigby2011} revealed a pronounced decline in the comoving space density of luminous radio sources above $z\!\sim\!3$--5, leading to the notion of a sharp ``cut-off'' in the RAGN population.  However, these studies are based on bright, low-frequency samples that probe only the most extreme objects and are subject to strong selection biases against compact or faint high-redshift systems.  As a result, it remains unclear whether this observed decline reflects genuine cosmic evolution or is merely a consequence of survey limitations.

Over the past few years, the discovery of several powerful radio AGN at $z>5.5$ (and out to $z\sim7$, see Section~\ref{sec:observations}) has opened a new window on the physical conditions governing SMBH growth during the first gigayear of cosmic history.  Yet these detections remain too scarce to constrain formation scenarios, and observations alone cannot determine whether such sources are exceptional outliers or the luminous end of a broader, still-undetected population.  Theoretical modelling is therefore indispensable for assessing the true prevalence of powerful accretion and jet activity in the early Universe, for developing robust diagnostics to identify them, and for predicting what the next generation of deep radio surveys, particularly those conducted with the SKAO, should reveal in terms of the earliest accreting SMBHs.  Cosmological simulations provide the natural framework to connect the observed radio AGN population with the physical processes that drive SMBH formation, growth, and feedback during the EoR, and to forecast their detectability with forthcoming radio surveys.

In this section we summarise the state-of-the-art theoretical predictions for RAGN at $z>6$, focusing and extending on the results of \citet{Amarantidis2019} and related works.  Although that study targeted both X-ray and radio emission from AGN in the EoR, its radio component provides the first systematic assessment of the expected radio luminosity function (RLF) for AGN in the EoR.

\subsection{Exploring simulations of the early Universe}

Predicting AGN activity during the EoR can rely on cosmological models that track the co-evolution of dark matter, gas, stars, and SMBHs across cosmic time.  Two broad families of approaches have been developed to address this problem: hydrodynamical simulations (HDSs) model galaxy formation by numerically solving the equations of gravity and hydrodynamics in a cosmological context, following the evolution of dark matter, gas, and stars across discrete resolution elements. Semi-analytical models (SAMs), on the other hand, approximate these physical processes with parametric prescriptions that are applied along the merger trees of dark matter halos. Each framework offers distinct advantages and limitations, and together they provide the foundation on which most current predictions for the early Universe are built.  By exploring different prescriptions for seeding, accretion, and feedback, these models delineate the range of plausible SMBH evolutionary paths at the highest redshifts and quantify the theoretical uncertainties that remain in predicting their observable properties.

Hydrodynamical simulations explicitly resolve the interplay between gas cooling, star formation, feedback, and accretion within evolving large-scale structures.  Prominent examples include \mbox{\textsc{Illustris/TNG}} \citep{Vogelsberger2014,Nelson2019}, \textsc{EAGLE} \citep{Schaye2015},
\textsc{Horizon-AGN} \citep{Dubois2014}, \textsc{SIMBA} \citep{Dave2019}, and \textsc{MassiveBlack-II} \citep{Khandai2015}. These simulations reproduce many observed galaxy properties and the SMBH--host scaling relations at low redshift, while implementing sub-grid models for black-hole seeding, accretion, and feedback.  Black holes are typically inserted in newly formed haloes once a threshold mass is reached, with initial seed masses of $10^{4}$--$10^{6}\,\rm M_\odot$. Accretion proceeds through Bondi-like or torque-limited prescriptions, and feedback is released as thermal or kinetic energy.  Their main limitations remain computational: restricted volumes make it difficult to capture the rarest, most massive halos, while finite resolution hampers the detailed modelling of gas inflows and small-scale accretion physics.

Semi-analytic models provide a complementary, computationally efficient route by applying analytic recipes for gas cooling, star formation, and SMBH growth to the hierarchical growth of dark-matter halos.  Models such as \textsc{L-Galaxies} \citep{Henriques2015}, \textsc{GALFORM} \citep{Lacey2016}, \textsc{Shark} \citep{Lagos2018}, and \textsc{Meraxes} \citep{Mutch2016} include explicit prescriptions for black-hole seeding, radiative and kinetic feedback, and duty cycles.  Their major strength lies in the ability to explore large cosmological volumes at relatively low computational cost, allowing wide parameter-space studies and statistical samples of the rarest, most luminous AGN -- albeit at the cost of relying on simplified, parameterised descriptions of baryonic processes.

\begin{figure}
    \centering
    \includegraphics[width=0.9\linewidth]{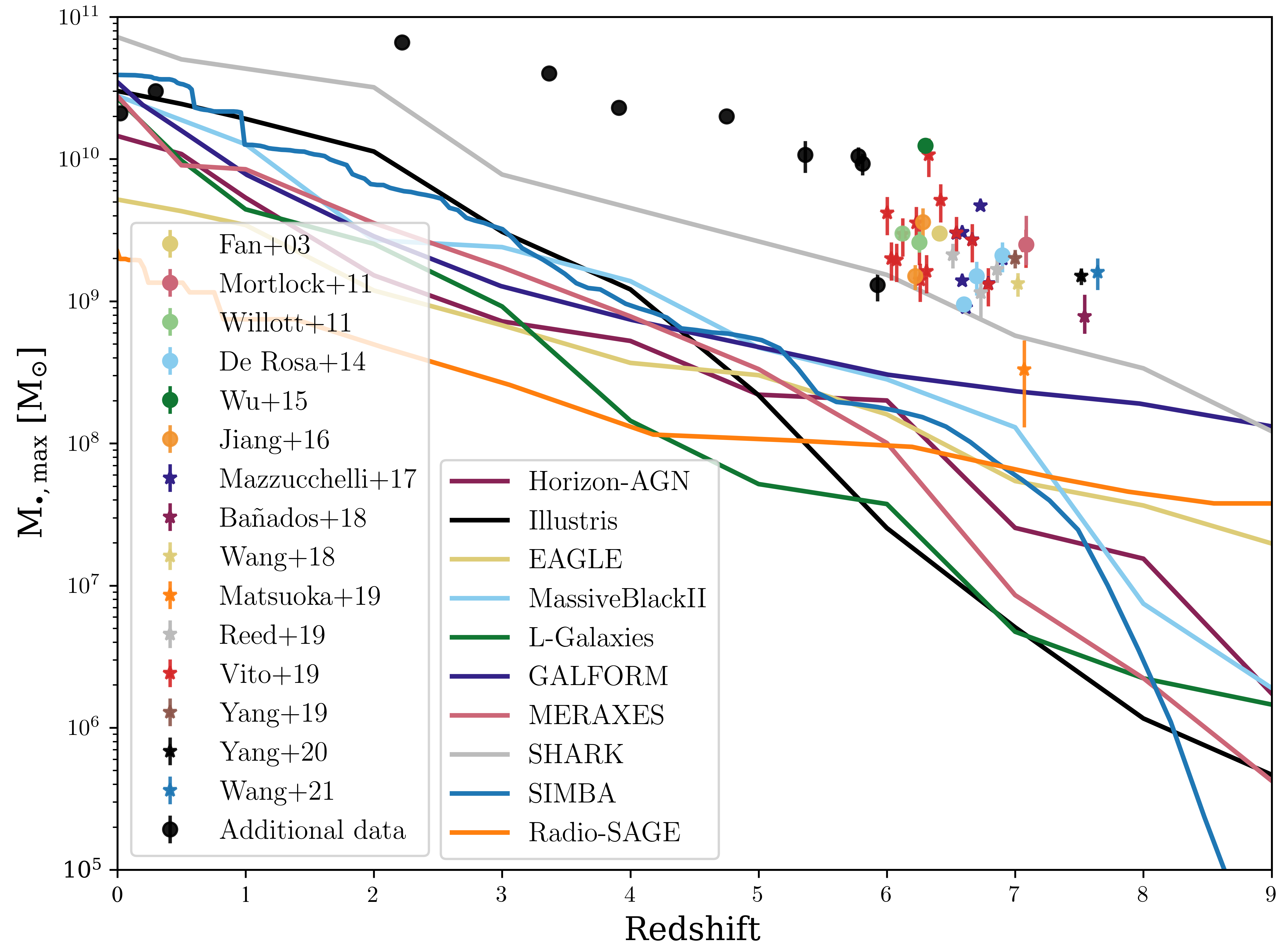}
    \caption{Maximum SMBH masses as a function of redshift for a range of cosmological simulations, extending the analysis of \citet{Amarantidis2019} to include more recent models, namely \textsc{SIMBA} and Radio-SAGE. Data points show observational estimates. Further discussion of the methodology and comparison datasets can be found in \citet{Amarantidis2019}.
    }
    \label{fig:MaxSMBHMass}
\end{figure}

These frameworks have been systematically compared only recently, as exemplified by \citet{Amarantidis2019}, who used a suite of cosmological models, spanning both HDSs and
SAMs, to explore the existence of RAGN populations in the EoR.  Their
analysis revealed
that all current models substantially underpredict the maximum SMBH masses compared
with those already observed, even at moderately high redshifts
\citep[Figure~\ref{fig:MaxSMBHMass}, see also][]{Habouzit2022}.  This limitation was interpreted as a consequence of
finite simulation volume, which restricts the sampling of the rarest, most massive
haloes hosting the brightest AGN.  \citet{Amarantidis2019} therefore argued that
predictions from current state-of-the-art galaxy models should be considered lower
limits to the true number of rare, high-luminosity AGN at very high redshift, and that
simply increasing the simulated volume would significantly improve agreement between
models and observations.  

In addition to these limitations, translating the SMBH properties predicted by
cosmological models into observable radio emission remains non-trivial.  Most
simulations do not include jets explicitly or compute their kinetic power
self-consistently.  Earlier post-processing approaches generally assumed a fixed
fraction of the accretion power to be converted into kinetic output, and then applied
empirical or semi-empirical relations to link jet power with radio luminosity
\citep[e.g.,][]{Willott1999,Cavagnolo2010,HeckmanBest2014}.  A more physically motivated treatment was developed by \citet{Fanidakis2011}, 
who coupled black-hole spin evolution and accretion physics within the \textsc{GALFORM} framework, to reproduce the radio luminosity scaling predicted by \citet{Heinz2003} and fit the observed local RLF. This model provides a physically grounded pathway from SMBH mass and accretion rate to observable radio emission, and underpins several subsequent implementations in cosmological post-processing studies.

Building on this framework, \citet{Amarantidis2019} applied such post-processing techniques to a range of HDSs and SAMs, providing one of the first systematic comparisons of predicted high-redshift AGN RLFs and revealing how differences in seeding, accretion, and feedback prescriptions drive large variations across state-of-the-art cosmological frameworks.

However, these approaches do not capture the dynamical effects of jet feedback on the surrounding gas.  More recent models have begun to address this limitation. The hydrodynamical simulation \textsc{SIMBA} \citep{Dave2019,Thomas2021} introduces a kinetic feedback mode in which low-Eddington accretion triggers bipolar, momentum-driven outflows that reproduce large-scale energetic effects of AGN jets, though without explicitly modelling their radiative (radio) emission.  A further advance was implemented in the SAM Radio-SAGE \citep{Raouf2017}, which combines the self-consistent growth and feedback of SMBHs with an analytic treatment of jet evolution and lobe expansion \citep{Turner2015}, allowing both mechanical energy injection into the halo gas and the prediction of observable radio luminosities.  

\begin{figure}[t]
    \centering
    \includegraphics[width=0.9\linewidth]{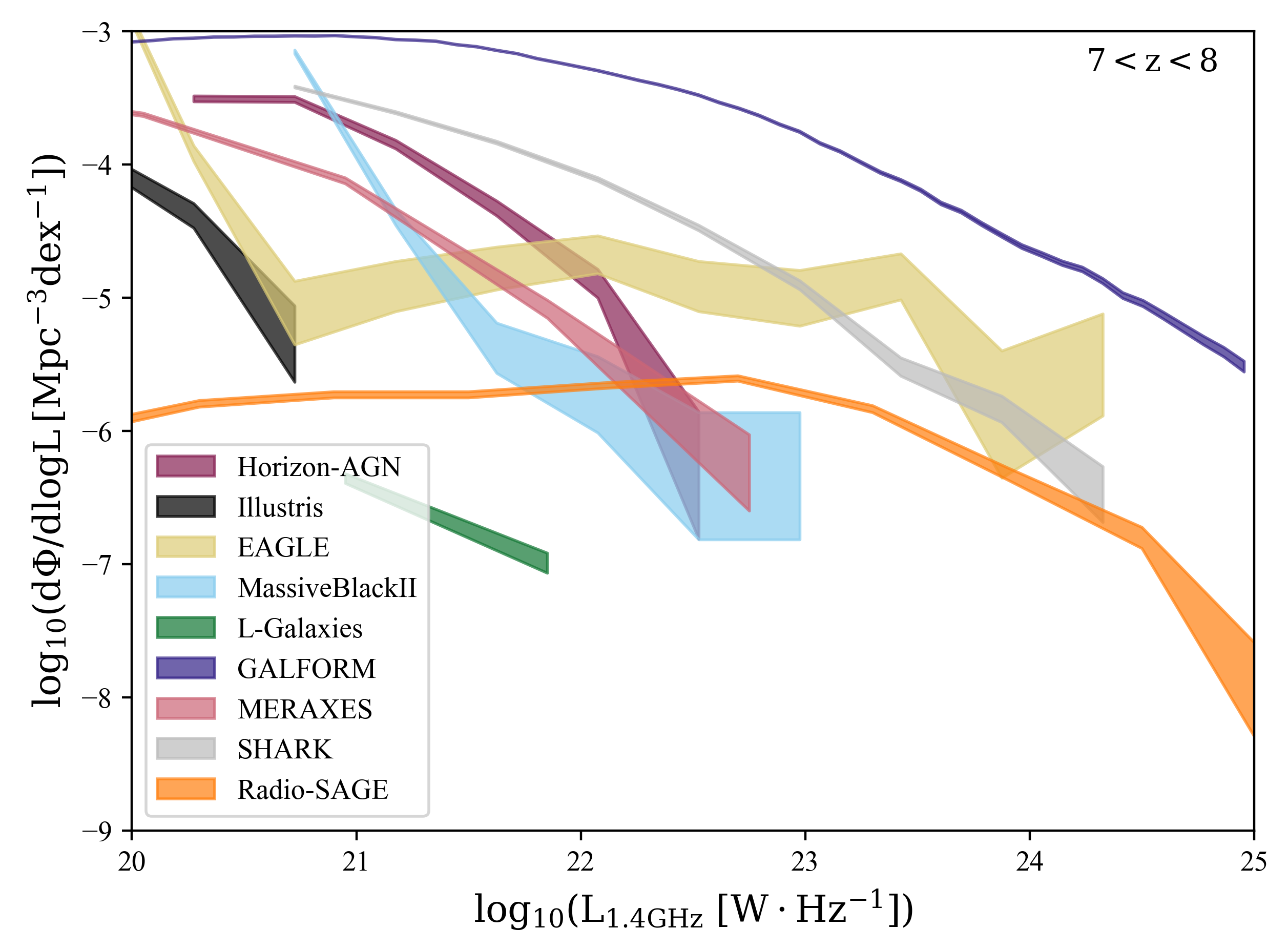}
    \caption{Predicted radio luminosity functions for AGN at $z\!\sim\!7$--$8$ derived from a range of cosmological simulations, extending the analysis of \citet{Amarantidis2019}.  Coloured lines correspond to different simulation frameworks, both hydrodynamical (Horizon-AGN, Illustris, EAGLE, MassiveBlackII) and semi-analytic (L-Galaxies, GALFORM, MERAXES, SHARK, Radio-SAGE).
    }
    \label{fig:RLF_predictions}
\end{figure}

Figure~\ref{fig:RLF_predictions} shows the predicted RLFs for AGN at $z\!\sim\!7$--$8$, originally derived by
\citet{Amarantidis2019} and extended here to include additional cosmological models and
updated scaling relations between SMBH accretion rates and radio output.  
When accounting for the fact that current simulations systematically underpredict the
most massive black holes -- those most closely associated with powerful radio emission
-- it becomes evident that state-of-the-art galaxy formation and evolution models do
support the existence of a substantial population of radio-powerful sources well into
the EoR.  The predicted space densities, though still subject to large model
uncertainties, suggest that a non-negligible fraction of these sources should be within
reach of forthcoming deep SKAO surveys.

Beyond volume and modelling effects, numerical resolution also plays a crucial but often neglected role. This effect can be particularly important at high redshifts \citep[e.g.,][]{Covas2025}, with lower-resolution simulations yielding systematically smaller SMBH masses and accretion rates than high-resolution runs, especially for black holes in haloes below $\sim10^{12}\,\rm M_\odot$, a critical regime for high-redshift predictions.  This again leads to an underrepresentation of luminous AGN and a suppressed bright end of the simulated luminosity functions, particularly serious at the highest redshifts.

In summary, current simulations and models provide a valuable but incomplete picture of early SMBH evolution.  They bracket the uncertainties associated with limited volume, resolution, and the treatment of accretion and feedback, but they do not yet converge on the abundance or maximum masses of black holes during the EoR.  Nevertheless, these limitations themselves imply that a substantial population of RAGN may remain undetected but within reach of upcoming (and even current) facilities.  Indeed, much of the most recent evidence from confirmed high-redshift radio sources appears consistent with this expectation, as we will discuss in Section~\ref{sec:observations}.  Further progress requires a more physically grounded treatment of jet production and radio emission in
cosmological models, an issue we address in Section~\ref{sec:optimisation}.

\section{Observational Constraints and Current Surveys}\label{sec:observations}

Understanding the nature and abundance of RAGN at the highest redshifts requires confronting theoretical predictions with the present observational landscape.  Over the past two decades, our view of the radio sky has been transformed by major advances in continuum sensitivity, bandwidth, and survey speed.  Facilities such as the Very Large Array (VLA), the Australia Telescope Compact Array (ATCA), the Westerbork Synthesis Radio Telescope (WSRT) and the Giant Metrewave Radio Telescope (GMRT) progressively expanded the accessible parameter space, and, more recently, SKAO pathfinder and precursor instruments have driven a qualitative change in our ability to map faint radio populations.  Wide-area surveys with LOFAR and the MWA now provide broad low-frequency coverage (50–350\,MHz), with LoTSS \citep{Shimwell2017} and GLEAM \citep{Wayth2015,Callingham2017} mapping hundreds of thousands of sources, while at higher frequencies ASKAP \citep[e.g., through EMU,][]{Hopkins2025} and MeerKAT (e.g., through MIGHTEE, \citealt{Jarvis2016}, and MeerKLASS,  \citealt{Santos2016}) routinely reach $\mu$Jy sensitivities over tens to thousands of square degrees.  Together these surveys probe the faint radio-emitting galaxy population in unprecedented detail, bridging the luminosity range between classical radio galaxies and normal star-forming systems.

In spite of this observational revolution, the direct detection and confirmation of RAGN in the EoR remains exceptionally challenging.  The number of securely identified radio sources beyond $z\!\sim\!6$ is still limited to a few dozen \citep[e.g.,][]{McGreer2006,Belladitta2020,Banados2021,Banados2025,Gloudemans2022,Endsley2023,Ighina2025}, and most have been discovered as optically selected quasars subsequently found to host radio emission \citep[e.g.,][]{Ighina2021,Gloudemans2021}.  This observational bias has restricted our view of early AGN activity to the brightest and most accessible systems, while the fainter, potentially more numerous radio population remains largely unconstrained.

In this section we review the current state of deep radio surveys and their
multi-wavelength counterparts, summarising the observational evidence for RAGN at high redshift and the methods used to identify them.  We briefly assess the efficiency and limitations of existing selection techniques, the biases that affect faint-source studies, and how these challenges compare with theoretical expectations discussed in Section~\ref{sec:theory}.  These considerations provide the foundation for the next section, where we explore how the SKAO will enable the development of more robust, physically motivated, radio-based diagnostics capable of revealing the earliest AGN in the Universe.

\subsection{Sampling the optically bright high-redshift RAGN population}

The vast majority of high-redshift RAGN currently known have been identified through their bright rest–frame ultraviolet (UV) emission. The advent of wide-area optical and NIR photometric surveys such as the Panoramic Survey Telescope and Rapid Response System (Pan-STARRS; \citealt{Chambers2016}), the Dark Energy Survey \citep[DES;][]{Abbott2021}, and the VISTA Kilo-degree Infrared Galaxy Survey \citep[VIKING;][]{Edge2013} has enabled the discovery of hundreds of AGN at $z>6$ 
\citep[e.g.,][]{Venemans2013,Banados2016,Reed2019}. Most high-$z$ AGN searches rely on the so-called Ly$\alpha$ dropout technique \citep[e.g.,][]{Fan2006,Banados2014,Wang2019a,Shobhana2023}, which selects objects with a strong colour break between two consecutive filters, one probing the continuum at rest-frame wavelengths above Ly$\alpha$ and the other sampling wavelengths below Ly$\alpha$, where the flux is almost completely absorbed by the IGM. Despite some contamination (e.g., from brown dwarf stars), this method has led to the discovery of more than 300 AGN at $z>5$ (see \citealt{Fan2023} for a recent review). Other approaches, including machine-learning algorithms, have been explored to improve selection efficiency, but they also predominantly identify UV-bright quasars \citep[e.g.,][]{Wagenveld2022,Ye2024}.

The radio detection of a subset of these AGN in legacy surveys such as the NRAO VLA Sky Survey \citep[NVSS;][]{Condon1998} and the Faint Images of the Radio Sky at Twenty-Centimeters \citep[FIRST;][]{Becker1994} at 1.4\,GHz enabled the identification of quasars hosting powerful radio jets already at the EoR \citep[e.g.,][]{McGreer2006, Willott2010}.  A major advance came with the new generation of wide-area radio surveys, including LoTSS \citep{Shimwell2017} at 144\,MHz, the Rapid ASKAP Continuum Survey \citep[RACS;][]{McConnell2020} at 0.888, 1.37 and 1.66\,GHz, and the Very Large Array Sky Survey \citep[VLASS;][]{Lacy2020} at 3\,GHz.  Their improved sensitivity, combined with dedicated searches for RAGN guided by optical/NIR quasar-selection criteria, has led both to the radio detection of known high-$z$ quasars \citep[e.g.,][]{Ighina2021,Gloudemans2021} and to the discovery of new ones \citep[e.g.,][]{Gloudemans2022,Ighina2023,Banados2025}.

Combined optical/NIR and radio studies have also led to the discovery of numerous high-$z$ Flat Spectrum Radio Quasars \citep[FSRQs, or blazars; see][]{Belladitta2020,Ighina2024,Banados2025}. These are AGN with relativistic jets oriented close to our line of sight ($\theta_{\rm view}\lesssim1/\Gamma$, where $\Gamma$ is the jet’s bulk Lorentz factor). By definition, this population should not suffer from obscuration, as the jet is expected to clear the line-of-sight. Relativistic beaming effects make blazars observationally distinct from the general jetted AGN population, showing a flat radio SED and compact (VLBI-scale) powerful radio emission \citep[e.g.,][]{Coppejans2016,Caccianiga2019,An2020}, together with flat X-ray spectra \citep{Ghisellini2014b,Ighina2019}. Importantly, given their geometry, the detection of a single blazar implies the presence of $N_{\rm tot}\approx2\Gamma^2$ jetted AGN with similar intrinsic properties and redshift but viewed at larger angles (with $\Gamma\sim5-10$; \citealt{Ghisellini2014b,Spingola2020}). Consequently, constructing well-defined blazar samples allows us to infer the total number density of jetted AGN in a given redshift range \citep[e.g.,][]{Diana2022,Sbarrato2022}. Following this approach, several independent studies \citep[e.g.,][]{Caccianiga2024,Banados2025} compared the space density of jetted AGN inferred from blazar counts with that derived from UV luminosity functions, finding that the jetted fraction may increase to $\sim80\%$ at $z\sim7$. If confirmed, such a high fraction would have major implications for models of SMBH growth in the early Universe. However, this result may also reflect an underestimated correction for obscured AGN in optical luminosity functions \citep[e.g.,][]{Caccianiga2024,Banados2025}. 

These efforts have rapidly expanded the number of known $z>6$ RAGN.  However, such selections still depend on bright optical/NIR counterparts to apply the Ly$\alpha$ dropout method and to confirm redshifts spectroscopically from the ground.  This dependence introduces strong observational biases that restrict our view to a few exceptional systems, leaving the RAGN population at high redshift sampled in a highly biased way.

\subsection{The elusive obscured population of high-redshift RAGN} \label{sec:elusiveHzRAGN}

While the vast majority of AGN currently known at high redshift are by selection optically unobscured, they likely represent only a small fraction of the total population. Several studies predict that obscured sources dominate the AGN census at early cosmic epochs, with more than 80\% expected to be heavily obscured at $z>5$ \citep[e.g.,][]{Gilli2022}.  This high fraction arises from large-scale obscuration by their host galaxies, which at early times are denser, dustier, and more clumpy \citep[see][]{Kartalpede2023}. Despite their expected prevalence, obscured AGN are difficult to identify because their optical/NIR emission is strongly attenuated. Mid-infrared (MIR) observations can in principle isolate obscured AGN through characteristic hot-dust colours and MIR excess. However, at higher redshifts dedicated MIR observations (namely with JWST) are needed, making this approach observationally expensive and practically infeasible for a large dataset of candidates. The use of radio-based selection criteria can help to overcome the biases inherent to optical/NIR searches and uncover the heavily obscured RAGN population at high redshift, bringing us closer to revealing the earliest stages of black-hole growth.

However, as discussed in Section~\ref{sec:Introduction}, spectral index alone appears to be insufficient to isolate high-redshift RAGN at the faint flux densities now reached by deep surveys, underscoring the need for more refined, multi-parameter approaches. Criteria based on \textit{spectral curvature} have emerged as a promising alternative to simple spectral-index selection \citep{Drouart2020,Broderick2022}.  By sampling the radio SED at three or more frequencies, it becomes possible to identify convex or peaked spectra associated with compact or young AGN whose spectral turnover shifts to lower observed frequencies at increasing redshift.  

\begin{figure}
\centering
\includegraphics[width=\linewidth]{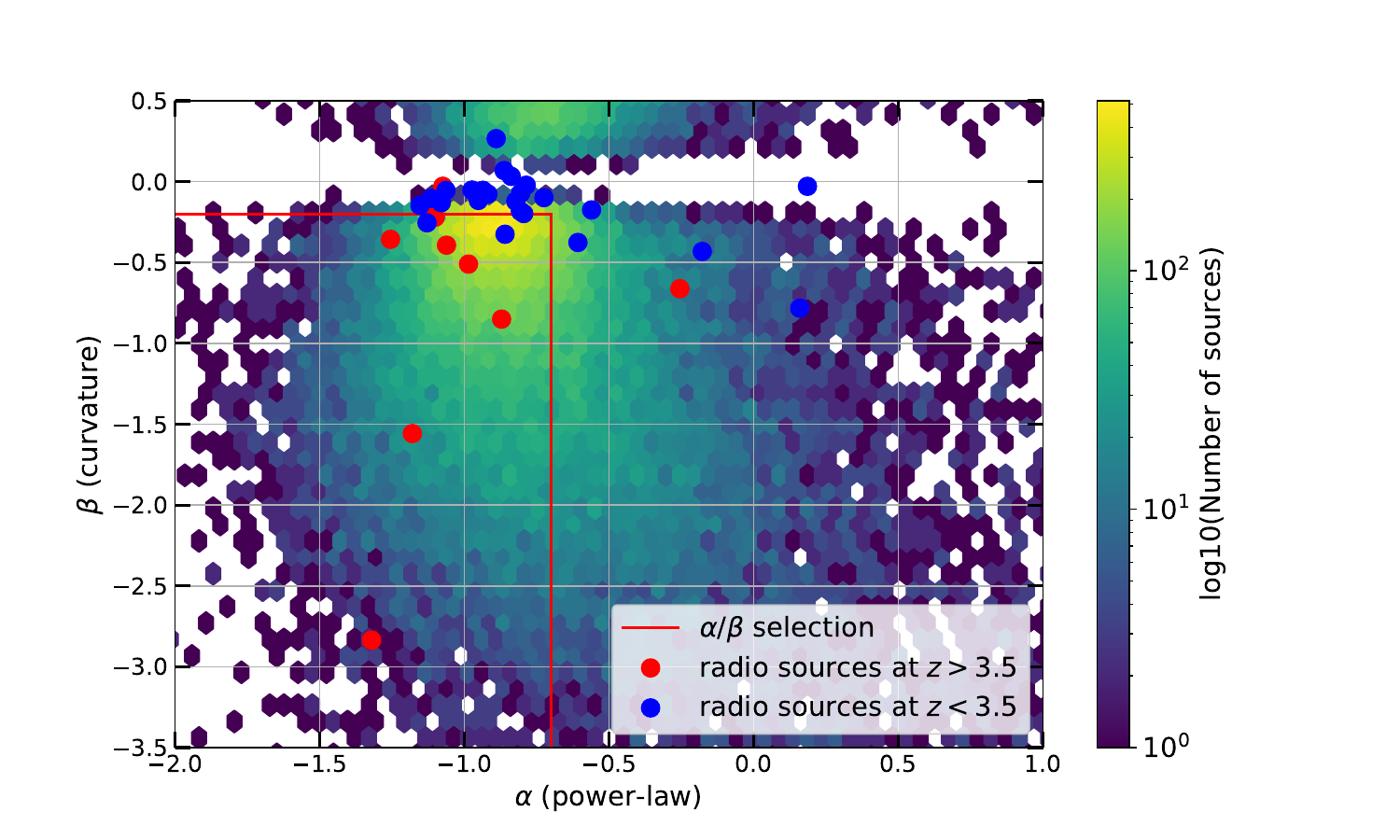}
\caption{
Spectral-index versus curvature parameter space for sources detected in the GLEAM survey with curved low-frequency radio spectra. Red and blue points mark known $z>2$ radio galaxies with $L_{\rm 500\,MHz} > 10^{27}\,\mathrm{W\,Hz^{-1}}$. Seven of the nine sources at $z>3.5$ lie within the region outlined by the red solid lines, which defines the selection space for identifying candidate high-redshift, powerful radio sources \citep[see][for details]{Broderick2022}.}
\label{fig:curvature}
\end{figure}

The unique broadband coverage (70-230\,MHz) of the MWA enables this approach to be applied systematically (see Figure~\ref{fig:curvature}). Using data from the GLEAM survey \citep{Wayth2015}, \citet{Drouart2020} discovered a radio-luminous galaxy at $z=5.55$ (GLEAM~J0856+0223) in a pilot study covering only 60~deg$^2$. This source was selected purely on the basis of its radio properties (spectral index and curvature), and its non-detection in the NIR VIKING survey. Comparable analyses in other surveys require assembling multi-band data from different facilities, a process still limited by inhomogeneous sensitivity and calibration.  As a result, only a modest number of sources can currently be evaluated through this method, though it remains one of the most physically motivated routes for identifying distant RAGN, as we will see in Section~\ref{sec:optimisation}.

An important complication for radio-based selection at very high redshift is the suppression of radio emission by inverse-Compton scattering of synchrotron electrons off the cosmic microwave background (CMB). This ``CMB muting'' becomes increasingly significant at high redshift (\mbox{$\propto (1+z)^4$}; e.g., \citealt{Ighina2021}), preferentially dimming the extended lobe emission and causing distant sources to appear more compact or even radio-weak despite hosting powerful relativistic jets \citep[e.g.,][]{Ghisellini2014a,Afonso2015}.  The resulting changes in spectral slopes complicate both flux-limited source selection and the interpretation of spectral curvature, further emphasising the need for physically motivated, broadband spectral modelling to reliably identify the most distant RAGN.

\subsection{A biased view of high-redshift RAGN}

Given the mixed and intrinsically biased view provided by existing selection criteria, the discovery of high-redshift RAGN has progressed in a notably non-uniform way. Progress has been gradual and strongly shaped by the differing selection methods employed across surveys.  The first radio-selected galaxy beyond $z=5$, TN\,J0924$-$2201 \citep{vanBreugel1999}, marked a key milestone, but it was followed by a long period during which most newly discovered $z>5.5$ active nuclei were identified through their optical properties rather than their radio emission.  Many of these optically selected quasars -- such as SDSS\,J0836+0054 at $z=5.82$ \citep{Fan2001}, CFHQS\,J1429+5447 at $z=6.21$ \citep{Willott2010}, and SDSS\,J2228+0110 at $z=5.95$ \citep{Zeimann2011} -- were subsequently found to host powerful radio emission with luminosities comparable to those of classical radio galaxies. 

More recently, dedicated radio-based searches have uncovered new high-redshift RAGN, progressively extending into the EoR, that were not pre-selected optically.  These include GLEAM J0856+0223 at $z=5.55$ \citep{Drouart2020}, VIK\,J2318$-$3113 at $z=6.44$ \citep{Ighina2021}, PSO J172.3556+18.7734 at $z=6.82$ \citep{Banados2021}, and the likely blazar J0410$-$0139 at $z=7.0$ \citep{Banados2025}.  These discoveries demonstrate that the physical conditions required to launch and sustain powerful radio jets were already in place within the first gigayear after the Big Bang \citep[see also][]{Gloudemans2022,Endsley2023}.  

\begin{figure}
\centering
\includegraphics[width=0.95\linewidth]{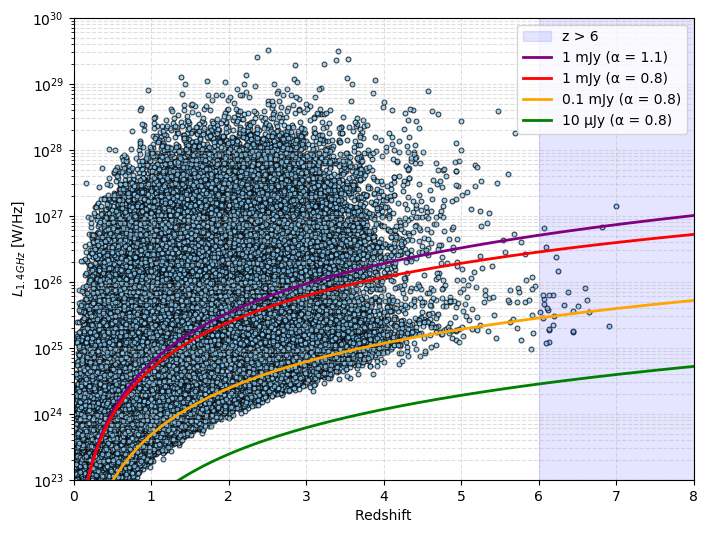}
\caption{A non-exhaustive census of known radio sources, showing radio luminosity as a function of redshift. Most objects at $z<5$ are drawn from the Million Quasars Catalogue \citep{Flesch2023}, cross-matched with LoTSS, RACS, and VLASS. The
heterogeneous distribution reflects the diversity of selection techniques and survey
sensitivities contributing to the current sample, illustrating the ongoing expansion of radio detections toward the earliest cosmic epochs. Coloured curves indicate the minimum detectable luminosity as a function of redshift for representative flux density thresholds and
radio spectral index $\alpha$ (defined as $S_\nu \propto \nu^{-\alpha}$). These limits illustrate the strong sensitivity bias shaping the observed population: while mJy-level surveys are restricted to the most radio-powerful AGN at the highest redshifts, sub-mJy and $\mu$Jy-depth observations dramatically expand the accessible luminosity range. This is particularly critical in the Epoch of Reionisation ($z \gtrsim 6$, shaded region), where a growing number of detections are being made at flux density levels of just a few tens of $\mu$Jy, including new radio-emitting QSOs recently revealed by the EMU survey (Caccianiga et al., submitted), underscoring the pivotal role of next-generation SKAO surveys in unveiling the earliest RAGN population.}
\label{fig:census}
\end{figure}

In Figure~\ref{fig:census} we show a non-exhaustive census of radio sources currently known out to the highest redshifts, identified through a variety of optical and radio selection techniques.  The distribution highlights the heterogeneous nature of existing samples but nevertheless makes clear that powerful radio emission was already present in the very early Universe.

Whilst the number of known RAGN in the EoR has grown to a few dozen, this remains far below the expectations from theoretical models (Section~\ref{sec:theory}).  The increasing rate of discovery of RAGN at $z>6$ strongly suggests the existence of a much larger population, very likely already present but still unidentified in wide radio surveys such as RACS. Upcoming wide-area NIR surveys such as those planned with ESA’s Euclid mission and  NASA’s Nancy Grace Roman Space Telescope will provide unprecedented opportunities to uncover these systems, covering $\sim$14,000\,deg$^{2}$ to $J\!\sim\!24.5$ \citep{Euclid2022} and $\sim$5,000\,deg$^{2}$ to $H\!\sim\!26.2$ \citep{Schlieder2024}, respectively. The synergy between these facilities and future SKAO surveys will enable an unparalleled census of the AGN population beyond $z>6$.

Nevertheless, a complete census of RAGN in the EoR will require a more radio-centric approach to source selection.  A physically motivated framework, able to exploit the full spectral information provided by forthcoming SKAO surveys, will be essential for identifying the earliest radio AGN reliably and systematically -- an issue we address in the following section.

\section{Developing Radio-only Diagnostics for Identifying the Earliest AGN} \label{sec:optimisation}

The limitations of empirical selection techniques discussed in the previous sections call for a fundamental change of paradigm. Physically motivated models of SMBH accretion, jet launching, and radio spectral evolution now provide the foundation for a new generation of radio-only diagnostics, and the sensitivity, continuous frequency coverage, and resolution of the SKA telescopes will, for the first time, provide the observational capabilities needed to realise this transition at the earliest cosmic epochs.

\subsection{Dynamical evolution of radio jets}

Dynamical models of radio jets \citep[see][for a review]{TurnerShabala2023} provide a framework for the required physically-motivated radio-only diagnostics. Such models describe how the initially relativistic AGN jets inflate radio-emitting lobes and continue to evolve throughout their active and remnant phases. Details of the jet-environment interaction sets the morphology of the resultant radio source, including whether it remains jetted or forms lobes or plumes; its particle composition, through jet mass-loading by entraining the interstellar medium (ISM) or stellar winds; and the resultant radio spectra. We refer the interested reader to \citet{TurnerShabala2023} and \citet{Turner2023} for details.

Observationally, size-luminosity (PD) tracks are the classical diagnostic used to characterise radio sources \citep{Baldwin1982,Hardcastle01.2026.SKA}. These can be used to study population properties including jet energetics, lifetimes and even mechanisms responsible for jet production \citep{Shabala2008,Turner2015,Hardcastle2019,Shabala2020}, while radio spectra or resolved brightness images provide additional leverage to measure the lobe magnetic-field strength and AGN duty cycle. In practice, however, estimating these quantities is non-trivial, as they typically require broadband radio coverage from low frequencies to the GHz regime to measure spectral curvature or break frequencies, together with source sizes, and they remain model dependent because the results depend on assumptions such as equipartition or minimum-energy conditions, as well as on the effects of particle mixing, spatially varying magnetic fields, and radiative losses \citep{Turner2018c,Turner2018b,Quici2025}.

On small galactic scales, jets propagate through the path of least resistance \citep[the so-called flood-and-channel phase of ][]{Sutherland2017}, imparting feedback to the multi-phase ISM. Once the jet plasma escapes this region, a jet breakout phase is observed, in which the jets expand essentially ballistically into the tenuous circumgalactic medium. Radio source morphology, on larger scales, depends sensitively on both jet and environment properties. If the initially conical jets run out of forward thrust before collimation, core-brightened Fanaroff-Riley type I (FR-I) jets are formed; while if collimation occurs edge-brightened Fanaroff-Riley type II (FR-II) sources are produced \citep{Krause2012}.

Radio emission is calculated in the models by tracking (in numerical models) or assuming (in analytical) sites of cosmic ray electron acceleration, then ageing these electrons due to adiabatic, synchrotron and inverse-Compton (off the CMB) losses. Because synchrotron and inverse Compton losses are proportional to the square of the Lorentz factor, high-frequency emission decays fastest, resulting in spectral steepening of optically thin spectra as the radio source ages; this forms the basis of spectral ageing analysis.

Analytical models are computationally efficient and hence essential for connecting theory to observations, but at least two effects require careful consideration, usually through numerical (magneto-) hydrodynamics which can then be applied to analytical models \citep[e.g.,][]{Turner2023}. First, turbulent mixing in forward- (in FR-Is) and back- (in FR-IIs) flowing material results in mixing of synchrotron-emitting cosmic ray electrons of different ages. Detailed calculations show that even a small fraction of recently accelerated electrons can significantly flatten the radio spectra at a given location in the lobe, explaining the observed discrepancy between dynamical and spectral ages in radio galaxies \citep{Turner2018a}. Second, inhomogeneities in the lobe magnetic field further bias spectral ages to younger estimates than the true values \citep{Hardcastle2013,Jerrim2025}.

The assumption of an optically thin plasma employed by models of FR-I/IIs may no longer be valid for compact AGNs. For radio emission on galaxy scales, free-free absorption by the hot and dense gas of the host galaxy has been shown to explain the observed low-frequency turnover  \citep{Bicknell2018,Young2025}, including the log-linear scaling of the turnover frequency with source size \citep{ODea1997,Orienti2014,Collier2018}. For the most compact jets, especially on VLBI scales, synchrotron self-absorption also plays an important role in setting the observed source spectra \citep{deVries2009}.

\subsection{The RAiSE framework: Bayesian inference of multiple radio jet parameters}

The most widely used current generation of analytical jet models are the \citet{Hardcastle2018} and the Radio AGN in Semi-analytic Environments \citep[RAiSE;][]{Turner2015,Turner2018c,Turner2018a,Turner2020,Turner2023}, dynamical models. Recently, a third analytical model by \citet{Beltran2025} has also been published. All of these provide a self-consistent description of the dynamical and radiative evolution of 10-100\,kpc-scale radio lobes within realistic cluster environments. 
While originally developed and calibrated at low redshift, these models are progressively being extended to cover the full range of effects relevant at high redshift, including ambient density evolution and inverse-Compton losses off the CMB, as well as the physical processes characteristic of compact, galaxy-scale systems such as interactions with an inhomogeneous interstellar medium, free-free absorption, and synchrotron self-absorption. Such developments make them particularly valuable for interpreting the radio properties of AGN during the EoR.

Critically for the task of creating a radio-only diagnostic, RAiSE aims to go beyond modelling integrated lobe sizes and luminosities, and also incorporate spatially resolved information. To do this, RAiSE explicitly models both jet and lobe-dominated phases of radio sources, as well as incorporating results from hydrodynamical simulations to inform model predictions. This enables RAiSE to produce realistic spatially resolved radio spectra, as well as capture the full evolution from the initial jet-dominated stage through the mature lobe-dominated phase, reproducing the transition from compact, young systems to extended FR-I and FR-II radio galaxies. The model naturally explains observed trends in surface brightness, spectral steepening, and morphology, producing complete, time-resolved radio SEDs from tens of MHz to tens of GHz. 

The rich set of synthetic observables produced by RAiSE is a key strength, enabling the inference of many intrinsic physical parameters -- jet kinetic power, source age, magnetic-field strength, and ambient density -- from observations in a Bayesian framework. The predictive capability of RAiSE makes it an ideal framework for developing the physically based diagnostics required in the SKAO era. By generating synthetic radio SEDs and evolutionary tracks for a wide range of jet powers, environments, and redshifts, the model provides the theoretical foundation for interpreting forthcoming broadband radio surveys and for identifying the earliest RAGN through their radio properties alone. The following sections present selected applications of RAiSE, illustrating how this framework can be used to interpret observations and guide the search for AGN activity during the EoR.

\subsection{Evolutionary tracks of powerful radio galaxies}
\label{sec:evoltracks}

The environments into which the jets expand, the energy density of the CMB inverse Compton photon field, and the relationship between rest frame and observed frequency all change appreciably with redshift. These effects result in redshift-dependent evolutionary tracks even for the same jet and environment parameters, therefore providing a potential probe of the observed source redshift, as we illustrate in this section.

\begin{figure}
\includegraphics[width=\textwidth, trim={89 0 68 38}, clip]{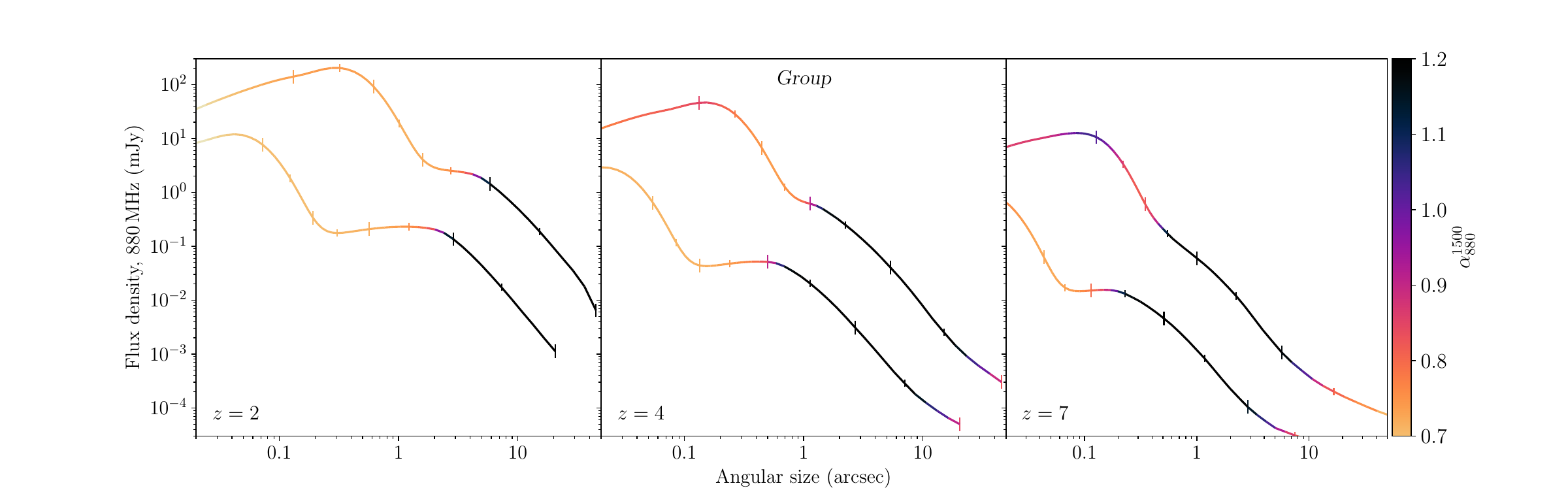}\vspace{6pt}
\includegraphics[width=\textwidth, trim={89 0 68 38}, clip]{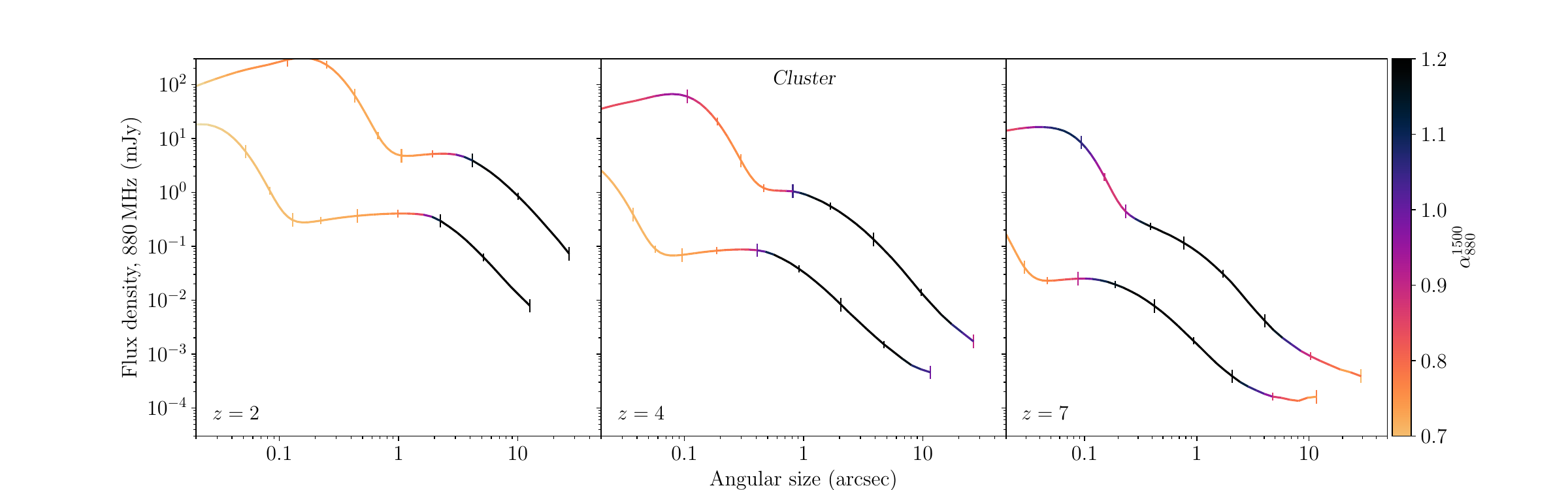}
\caption{RAiSE evolutionary tracks at 880\,MHz (observer frame) for moderate ($10^{37}$ W, bottom curves) and high-power ($10^{38}$ W, top curves) jets in group (top row) and cluster (bottom row) environments, at redshifts $z = 2$, $4$ and $7$. Colours indicate spectral index between 880 and 1500 MHz, and tick marks show log-linearly spaced source ages at $10^{-2}$, $10^{-1.5}$, $10^{-1}$ Myr, and so on, up to $100$ Myr.}
\label{fig:PD}
\end{figure}

In Figure~\ref{fig:PD}, we show the 880\,MHz evolutionary tracks in observed space (i.e. flux density and angular size) for two sources, broadly representative of moderate and high-power jets, evolved in $10^{12}$~M$_\odot$ (``group'') and $10^{13}$~M$_\odot$ (``cluster'') haloes, at a range of redshifts. These sources will be observable with the SKAO at all redshifts, at least in their compact ($\sim 1$ arcsec) phase. There is a general shift to lower flux densities and steeper spectra at high redshift, but the trends are complex; dynamical models are required to unravel the highly non-linear relationships between sets of observables in order to construct a radio-only diagnostic.

The interaction of the jet within the galaxy and its immediate surroundings adds further complexity. For instance, due to limited backflow and subsequent older lobe plasma ageing essentially as a remnant, the presence of the jet breakout phase can result in spectra which resemble restarted sources, as seen in simulations by \citet{Young2025}. High-resolution (sub-kpc) observations can distinguish between these scenarios, and of course provide tighter constraints on the inferred parameters for kpc-scale sources through a robust size measurement.

\subsection{Signatures of galaxy-scale jets}

\begin{figure}
\includegraphics[width=\textwidth, trim={89 0 114 42}, clip]{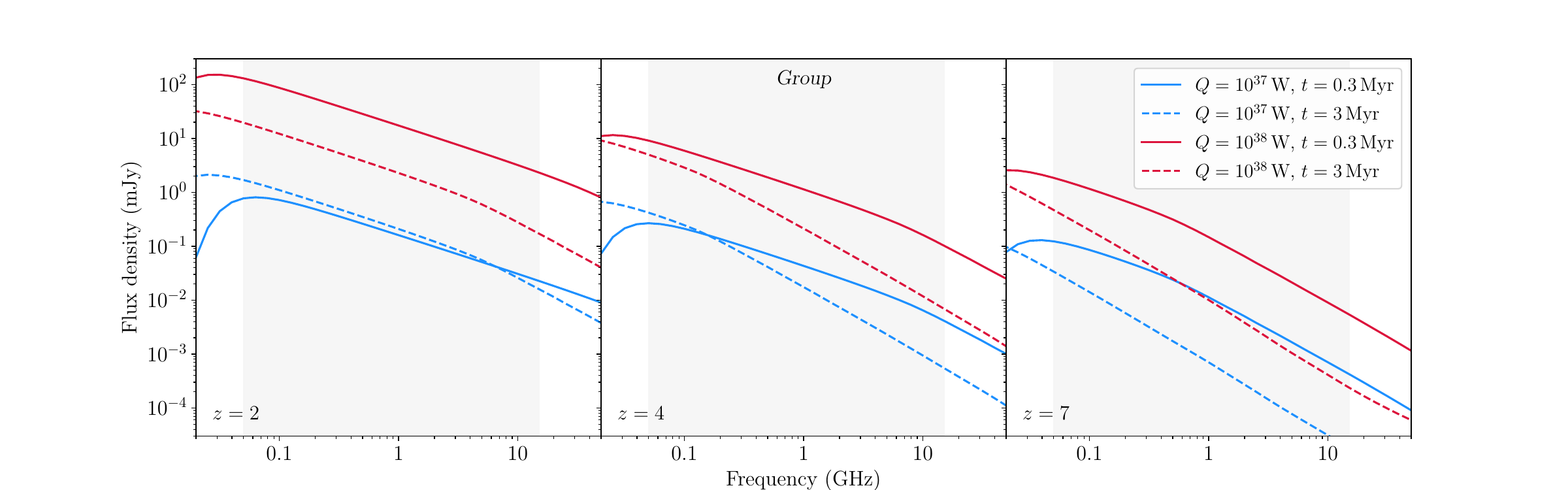}\vspace{6pt}
\includegraphics[width=\textwidth, trim={89 0 114 42}, clip]{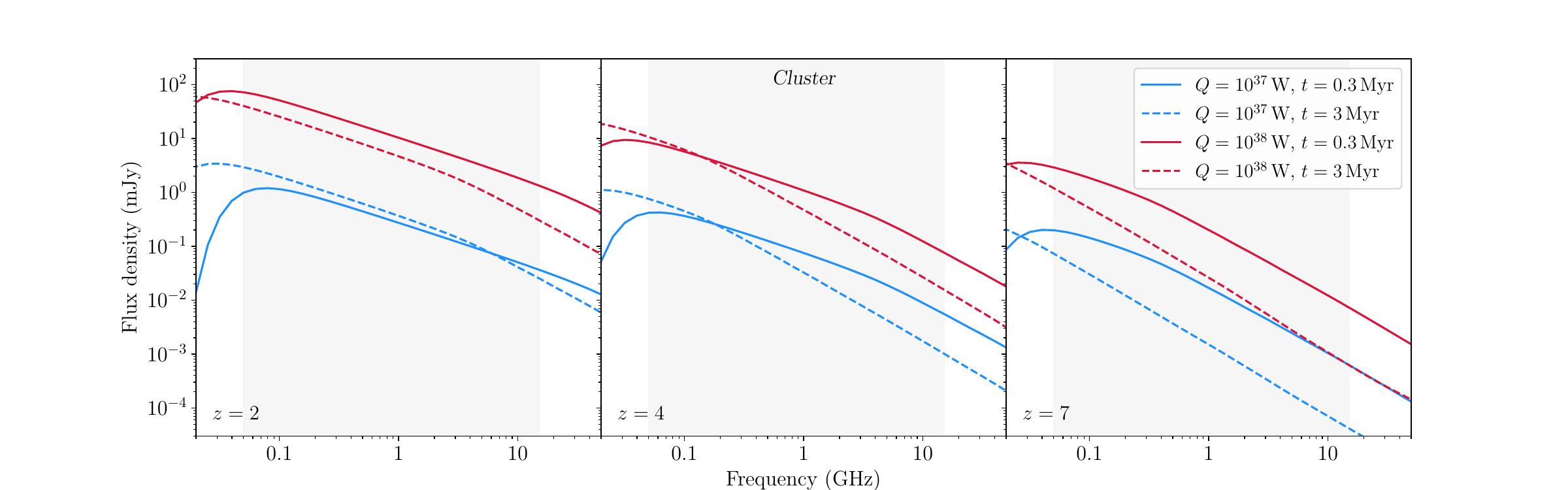}
\caption{Broadband radio SEDs for sources (as in Figure~\ref{fig:PD}) at different evolutionary stages: 0.3 and 3~Myr. The high-frequency spectral break is due to optically thin ageing. The low-frequency break is due to free-free absorption. Sampling the full shape of broadband radio SEDs provides useful constraints for estimating source redshifts.}
\label{fig:PD_spectra}
\end{figure}

The radio spectra (Figure~\ref{fig:PD_spectra}) for the moderate- and high-power radio galaxies considered in Section~\ref{sec:evoltracks} show that, at $z=7$, the optically thin break lies below 50 MHz for all but the youngest sources. Observable spectral curvature necessitates the use of intermediate-sized sources ($\sim 10$ kpc), which have higher rest-frame break frequencies than the larger classical double radio galaxies; typically these will be no more than a few Myr old. Such sources are also not large enough to suffer catastrophic inverse-Compton losses, despite the high redshift. 

At even smaller sizes, free-free absorption -- a process in which photons are absorbed by electrons in the presence of ions -- becomes significant for jets still confined to their host galaxies. In such cases, presence of a free-free turnover provides an additional constraint on the source redshift. There are now two critical frequencies: the optically thin break frequency discussed above, which relates to spectral ageing; and the optically thick turnover, which is determined by the amount of free-free absorbing material between the radio source and observer.

Hydrodynamic simulations of ISM-jet interactions in radio sources by \citet{Bicknell2018} and \citet{Young2025} have shown that, while ISM density plays a role in setting the normalisation of the optically thick turnover frequency--size ($\nu_{\rm ffa}$--$D$) relation, the evolutionary tracks in ($\nu_{\rm ffa}$--$D$) space are insensitive to jet power. By far the most significant factors are (i) source size, since the jets clear out the absorbing material as they burrow through the galactic gas; and (ii) the redshift of the source, which shifts the observed frequency of the turnover. The redshift dependence of source size is via the angular diameter distance; while at a fixed size, $\nu_{\rm ffa}$ is proportional to $(1+z)^{-1}$. Free-free absorbed sources with (likely VLBI) angular size measurements therefore provide another plausible avenue to estimate redshifts from radio continuum data; this will not be needed for sources which are resolved with the SKAO, which will also be too large to suffer significant free-free absorption.

\subsection{Redshifts from radio continuum data}

\citet{Turner2020redshifts} introduced a viable approach to estimate AGN redshifts based only on broadband radio continuum data, and showed that this approach works for powerful classical double sources expanding primarily within their halo environment -- such angular sizes are readily probed by the resolution of the SKAO. Their approach uses the physics encoded in dynamical models, observed radio source spectra and sizes (measured at a single frequency), and the divergence of luminosity and angular diameter distance with redshift, to find the best-fitting redshift solutions for each source. The dynamical model is calibrated to a small set of radio sources with known spectroscopic redshifts to reduce systematic uncertainties. 

\begin{figure}
\includegraphics[width=0.5\textwidth, trim={0 0 0 0}, clip]{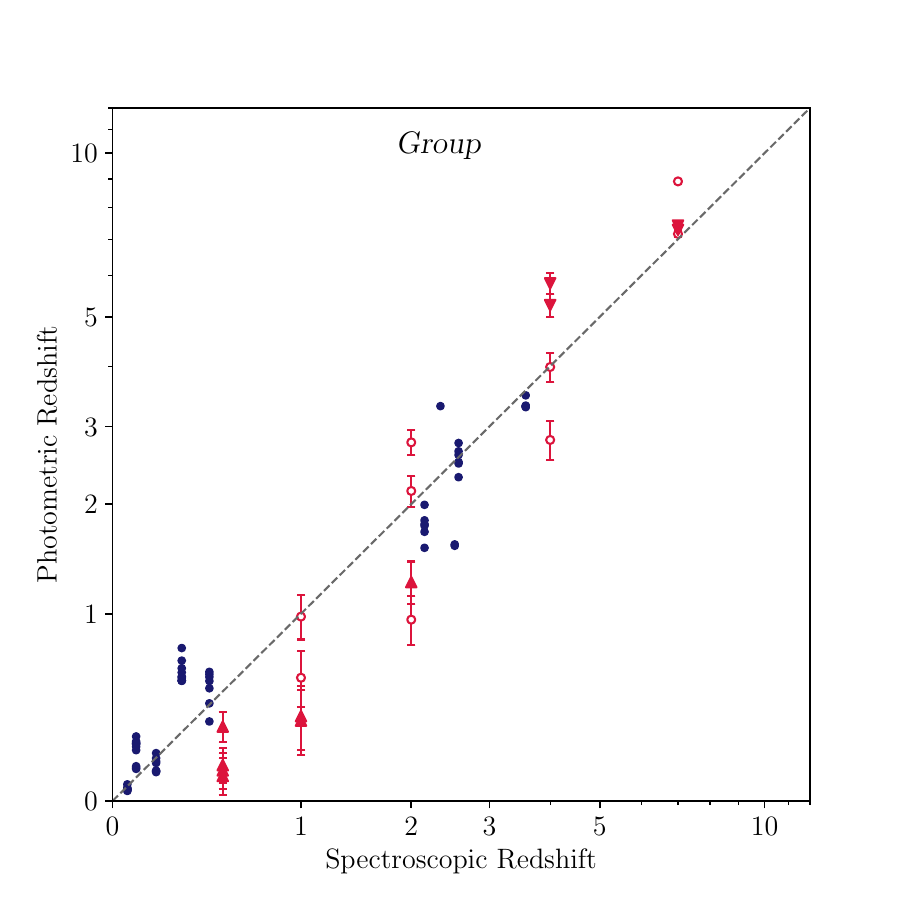}\includegraphics[width=0.5\textwidth, trim={0 0 0 0}, clip]{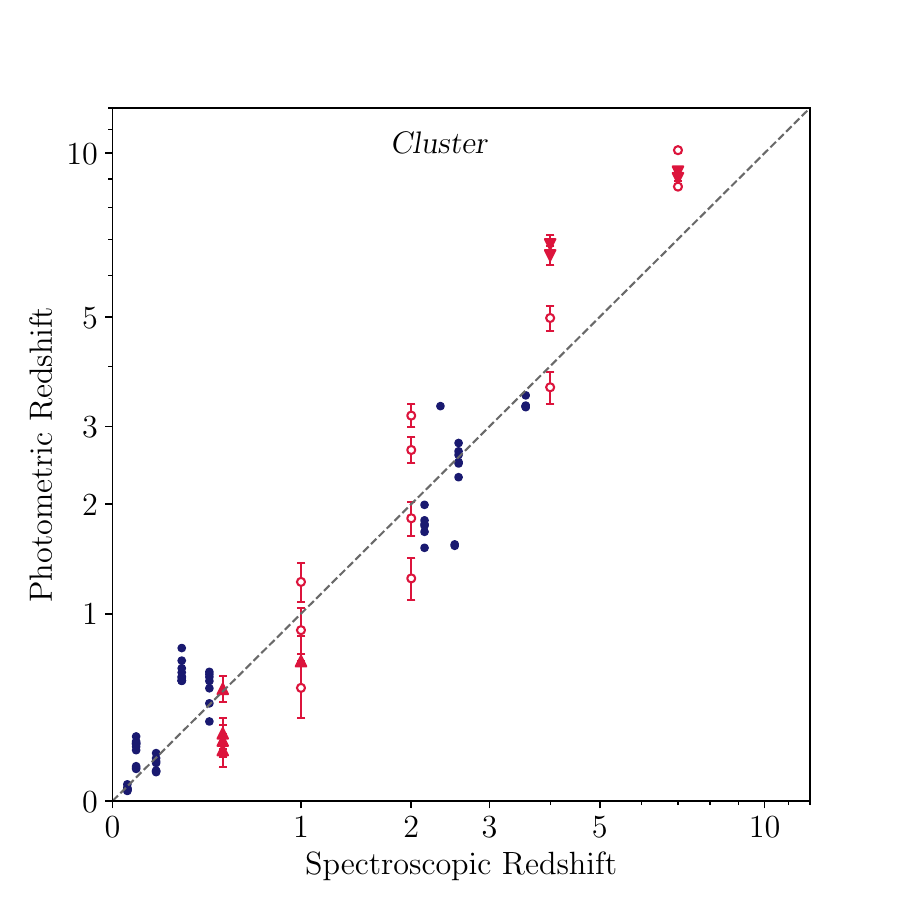}
\caption{Blue points: RAiSEred radio continuum redshift estimates for Cygnus A and fourteen higher redshift sources as a function of their spectroscopic redshift; the repeated points (for each spectroscopic redshift) represent different calibration data.
Red points: RAiSEred radio continuum redshift estimates for simulated lobed sources with the same jet powers, ages and halo masses as in Figure \ref{fig:PD_spectra}. The sources with a optically thin break frequency in the 50~MHz-15~GHz observing range of the SKAO are shown with unfilled red circles; those outside of the observing range with filled red circles.}
\label{fig:raisered}
\end{figure}

Figure~\ref{fig:raisered} shows the success of this `RAiSEred' fitting to Cygnus A and a subsample of radio sources from the 3CRR survey \citep{Laing1983} and the Herschel Radio Galaxy Evolution \citep[HeRGE;][]{Drouart2014} project (blue points; both panels).
The predictive power of the method is shown for simulated sources in two different environments up to $z=7$. In these plots, the error bars indicate uncertainties in the (radio) photometric redshift solution; these are largest at intermediate redshifts, where the angular diameter distance evolves little with redshift and therefore provides relatively poor constraints. These results suggest that such redshift calculations will be more useful at high ($z>5$) redshifts than at the ones at which most currently known classical doubles lie. The SKAO's sensitivity, broadband coverage and resolution will be sufficient to detect typical sources of this type at high redshifts (see Figure~\ref{fig:PD}).

\subsection{Probing early black hole assembly}

The strong evolution of radio luminosity and spectra over a jet lifetime, as described above, necessitates a revision of predictions for SMBH populations observable with the SKAO in the cosmological galaxy formation models outlined in Section~\ref{sec:theory}. 

\begin{figure}
\centering
\includegraphics[width=0.5\textwidth, trim={0 0 0 0}, clip]{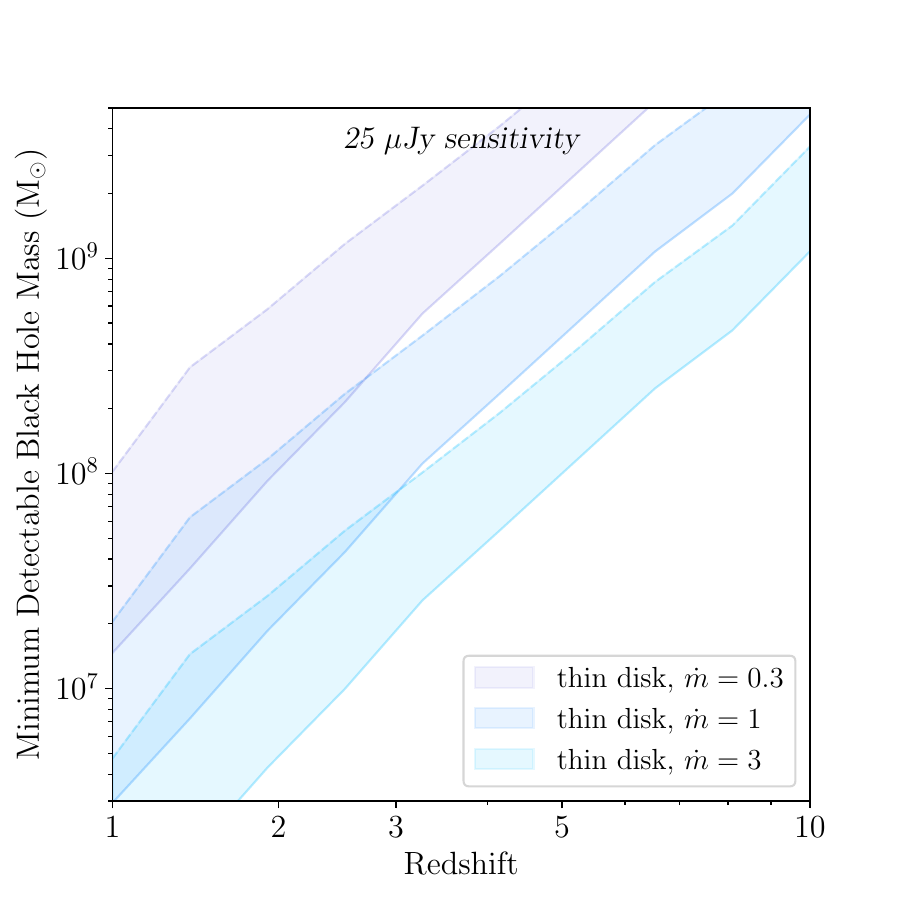}\includegraphics[width=0.5\textwidth, trim={0 0 0 0}, clip]{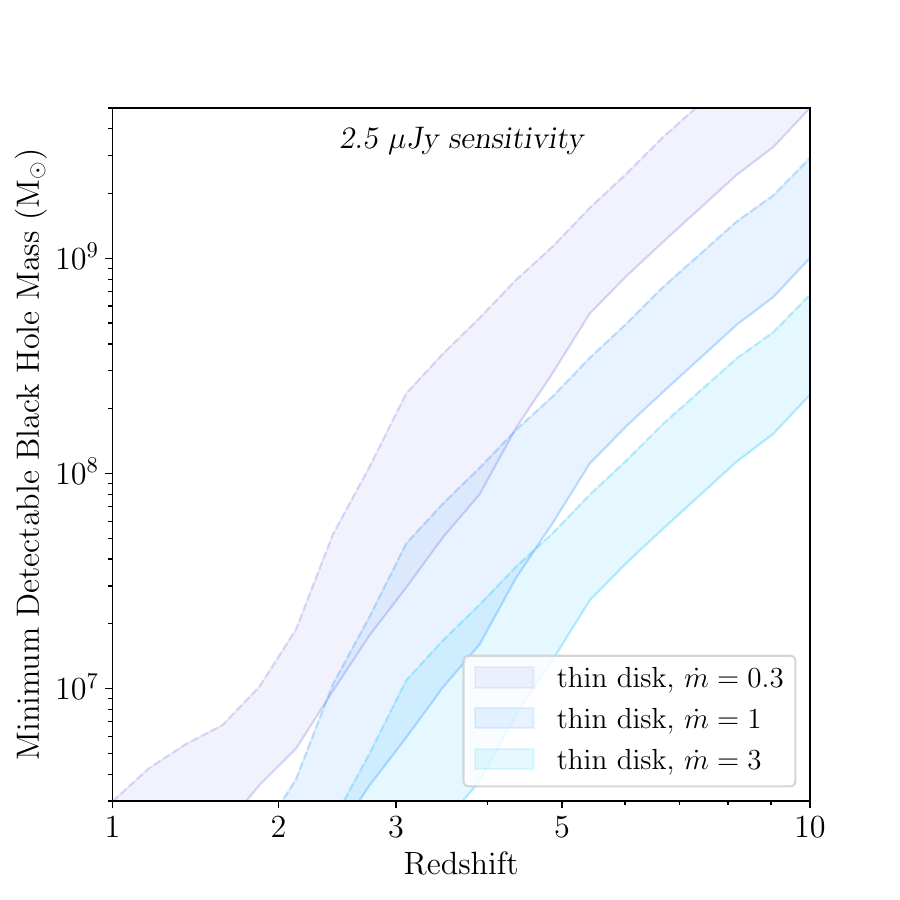}
\caption{Minimum detectable black hole masses as a function of redshift for radio galaxies observed in the plane of the sky. Predicted radio emission is for unbeamed jets and lobes, i.e. no relativistic beaming is included in the models. Models required a resolved source of angular size $>0.5^{\prime\prime}$ 
and an integrated flux density at 880\,MHz of at least 25\,$\mu$Jy (left) or 2.5\,$\mu$Jy (right). The conversion from jet power to black hole mass assumes a thin disk accretion flow with three plausible Eddington-scaled accretion rates: $\dot{m} = 0.3$, 1 and 3 (blue shading).}
\label{fig:Qmin_vs_z}
\end{figure}

To address this question, we use the RAiSE model to estimate the minimum jet kinetic power that yields a resolved source detectable by the SKAO at each redshift. 
For each jet power in the model, the maximum surface brightness is evaluated at 880\,MHz for sources larger than $0.5^{\prime\prime}$; for our assumed environments, at high redshifts this typically corresponds to intermediate source sizes slightly larger that the host galaxy, before inverse Compton losses become significant (cf. Figure~\ref{fig:PD}). We attempt to estimate the likely black hole masses by considering a conversion from jet power to SMBH mass. Because such a conversion is highly uncertain, we consider a plausible range of jet production efficiencies for the classical radiatively efficient Shakura-Sunyaev thin disks \citep{Meier2001}; this accretion flow is expected to be dominant mode in the EoR \citep{Amarantidis2019}. Figure~\ref{fig:Qmin_vs_z} suggests that black holes with mass as low as $10^{8.5}$~M$_\odot$ (25~$\mu$Jy sensitivity) or $10^8$~M$_\odot$ (2.5~$\mu$Jy sensitivity) may be detected in the EoR for mildly super-Eddington accretion rates (e.g., $\dot{m} \sim 3$).

We conclude with a brief discussion of model limitations and next steps. Environments around radio jets are a major source of uncertainty in all models. While expectations from cosmological models \citep[as in RAiSE,][]{Turner2015} or empirical relations \citep[as in][]{Hardcastle2018} provide a useful way forward, both these approaches become increasingly uncertain at higher redshifts. Direct measurements of jet environments through ancillary multi-wavelength data will be important in the future. Accurate interpretation of radio SEDs requires detailed consideration of both the complex jet dynamics \citep{Yates-Jones2023} and magnetic field structure \citep{Hardcastle2013,Jerrim2025}; integration of insights from detailed magnetohydrodynamic simulations into analytic models \citep{Turner2023} is an essential step for recovering the physical parameters (jet powers, ages) of the radio sources, as well as their redshifts. Finally, at low rest-frame frequencies both free-free absorption and synchrotron self-absorption effects will be important.

Notwithstanding these caveats, the integration of radio jet models with cosmological galaxy formation simulations and sensitive broadband data from the SKAO will provide groundbreaking advances into our understanding of the origin and evolution of the earliest SMBH populations.

\section{The Square Kilometre Array Observatory: Capabilities and Opportunity}
\label{sec:ska}

The SKAO will transform our ability to identify and characterise RAGN in the early Universe. In its AA4 configuration, the SKAO will provide broad frequency coverage spanning SKA-Low (50\,MHz -- 350\,MHz) and SKA-Mid (350\,MHz -- 15\,GHz), sub-arcsecond resolution across the SKA-Mid frequency range, and sensitivities reaching into the sub-$\mu$Jy regime. SKA-Mid will deliver sensitivities below 0.1\,$\mu$Jy at 1--2 GHz for deep integrations, while SKA-Low will reach tens of $\mu$Jy at 100 MHz with arcsecond-scale resolution \citep{Prandoni01.2026.SKA}. Together, these capabilities directly address the limitations highlighted in the previous sections: limited spectral sampling, insufficient depth at the faint end, and incomplete radio-SED characterisation. Most importantly, the combination of high sensitivity and broad instantaneous bandwidths will make it possible to sample the radio SED of each source with unprecedented detail, covering the full range of frequencies where synchrotron emission, spectral breaks, and curvature carry the imprint of jet physics (Section~\ref{sec:optimisation}), finally allowing to overcome the strong dependence on optical/NIR observations in the selection of very high redshift RAGN.

\subsection{Broadband radio SEDs: Tracing spectral shapes with SKAO sub-bands}

A key requirement for identifying high-redshift RAGN {\it from radio emission alone} is the ability to trace the full radio SED with sufficient frequency resolution. As shown in Section~\ref{sec:optimisation}, the precise shape of the radio spectrum -- its low-frequency turnover, spectral curvature, and high-frequency ageing break -- encodes essential information about jet power, age, environment, and redshift. Existing facilities can access parts of this information, and the MWA has already demonstrated
the power of dense low-frequency sampling for identifying distant candidates (e.g., via spectral-curvature selection, Section~\ref{sec:elusiveHzRAGN}). However, these capabilities remain limited in sensitivity, resolution, and frequency coverage. The SKAO will overcome these limitations by probing orders of magnitude deeper while sampling the radio emission across the entire $\sim$50\,MHz–15\,GHz range, enabling radio-only diagnostics capable of selecting distant RAGN even without optical counterparts.

\subsection{A SKAO continuum survey targeting the EoR}

Detecting the earliest radio-powerful SMBHs requires achieving sufficient sensitivity across the full SKAO frequency range to detect moderate-power $z\gtrsim7$ sources (Figure~\ref{fig:census}). For a conservative target luminosity of
$L_{1.4\,\mathrm{GHz}} = 4\times10^{24}\,\mathrm{W\,Hz^{-1}}$, the expected flux
density at $z\sim7$ is $\sim 10~\mu\mathrm{Jy}$ at 1.4\,GHz. A $5\sigma$ detection therefore requires a sensitivity of
$\sim 2~\mu\mathrm{Jy}$.

To reconstruct the full Band~2 SED, we assume ten equal-width sub-bands across the full bandwidth. Using Briggs weighting (robust parameter $=0$, slightly favouring sensitivity over angular resolution), the SKAO Sensitivity Calculator (AA4 configuration) indicates that $\sim4$ hours on-source per pointing are sufficient to reach the required $\sim2~\mu\mathrm{Jy}$ rms sensitivity in each sub-band, at an angular resolution of $\sim0.6^{\prime\prime} - 1^{\prime\prime}$ across the ten sub-bands. Adopting a noise effective area of 0.55\,deg$^{2}$ \citep{Prandoni01.2026.SKA} per pointing, a 20\,deg$^{2}$ Band~2 survey would require $\sim 150$\,hours. 

Adopting a spectral index of $\alpha=0.8$ ($S_\nu \propto \nu^{-\alpha}$) over the Band~1 to Band~2 frequency range, SKA-Mid Band~1 observations would need to reach matched sensitivities (rms) of $\sim3~\mu\mathrm{Jy}$ across its sub-bands. Adopting again 10 sub-bands and weighting scheme as before, approximately 7 hours per pointing would be required. The same 20\,deg$^2$ field could thus be covered by 14 pointings, totalling around 100 hours. 

For SKA-Low, achieving sufficient sensitivity to detect spectral turnovers while avoiding the confusion noise limit requires a two-pronged strategy within a single observation, designed to cover two complementary scenarios. A deep, broadband ($\sim$100\,MHz) image centred at 200\,MHz, using an optimised Briggs weighting and robust parameter to balance sensitivity against confusion, can reach $\sim4.5~\mu\mathrm{Jy}$ rms in $\sim$25 hours per pointing (depending on sky position and elevation constraints), remaining above the continuum confusion noise floor. A non-detection at this depth would itself be informative, placing a robust upper limit on the low-frequency flux density that can be compared against the higher-frequency detections to constrain the spectral shape and identify possible curvature or a low-frequency turnover. Conversely, if the source is detected, the same dataset can be imaged using 3--4 sub-bands within the 300\,MHz instantaneous bandwidth of SKA-Low, reaching $\sim$10--20\,$\mu\mathrm{Jy}$ rms per sub-band. This requires the exploration of more extreme robust imaging values, favouring angular resolution over sensitivity to mitigate confusion, a need that is most stringent at the lowest frequencies and progressively relaxes towards the upper end of the band. This sub-band imaging would then allow the low-frequency spectral shape to be mapped in detail. This imaging strategy, applied to the same survey data, provides both a robust detection/non-detection threshold and the means to map the spectral curvature when it is present. Covering the same 20\,deg$^{2}$ field as indicated before, this strategy would require a total of $\sim$8 pointings, hence $\sim$200 hours, driven by the smaller field of view at the higher frequencies of SKA-Low.

We note that, particularly at the lowest frequencies, careful consideration should be given to the selection of the observing field, as the associated observing conditions (e.g., elevation at transit, observations limited to periods close to transit) will have a substantial impact on the resulting sensitivity achievable. The possible presence of radio-frequency interference (RFI) can further limit the available bandwidth and sensitivity, and should also be accounted for when optimising the observing strategy.

At higher frequencies (Band~5a and Band~5b), the field of view is too small to support wide-area surveys. These bands should therefore be reserved for follow-up observations of the more promising high-redshift candidates identified across the 50\,MHz--1.8\,GHz range, providing essential constraints on spectral steepening and the high-frequency ageing break.

\subsection{Towards a complete RAGN census in the EoR}

By combining deep sensitivity, wide area, and full-band spectral characterisation, the SKAO will dramatically expand the known RAGN population at $z>6$. Its synergy with  Euclid and Roman (Section~\ref{sec:observations}) will provide NIR identifications for sources with faint or obscured optical emission, while SKAO-only detections will reveal the most heavily obscured systems.

The SKAO will therefore overcome the major limitations identified in earlier sections: optical biases (Section~\ref{sec:observations}), limited spectral sampling (Section~\ref{sec:optimisation}), and the underpredicted high-$z$ RAGN populations in simulations (Section~\ref{sec:theory}), leading to the first complete, physically grounded inventory of RAGN in the EoR.

\section*{Acknowledgements}

J.A., S.A., N.C., P.M., D.B., I.M. and C.P. acknowledge financial support from the Science and Technology Foundation (FCT, Portugal) through research grant UID/04434/2025 (DOI: 10.54499/ UID/04434/2025). 
The use of AI tools in this work was limited to supporting text editing and improving the readability of the manuscript.

\bibliographystyle{abbrvnat-maxbibnames4}
\bibliography{Afonso01_chapter} 

\end{document}